\documentclass[final,5p,times,twocolumn]{elsarticle}
\usepackage{graphicx}
\usepackage{amssymb}
\usepackage{amsmath}
\usepackage{lineno,hyperref}
\usepackage{color}
\hypersetup{colorlinks=true,linkcolor=red,citecolor=cyan,}
\modulolinenumbers[200]

\journal{Phys. Lett. B}

\begin{document}

\begin{frontmatter}

\title{Gravitational synchrotron radiation and Penrose process in STVG theory}

\author[addr1,addr2,addr3,addr4]
{Bobur Turimov\corref{cor1}}\cortext[cor1]{Corresponding author}
\ead{bturimov@astrin.uz}
\author[addr1,addr5]{Husan Alibekov}\ead{alibekov@astrin.uz}
\author[addr1,addr4]{Pulat Tadjimuratov}\ead{tajimura@astrin.uz}
\author[addr1,addr4,addr5,addr6]{Ahmadjon Abdujabbarov}
\ead{ahmadjon@astrin.uz}

\address[addr1]{Ulugh Beg Astronomical Institute, Astronomy St. 33, Tashkent 100052, Uzbekistan}\address[addr2]{School of Engineering, Akfa University, Milliy Bog St. 264, Tashkent 111221, Uzbekistan}\address[addr3]{Department of Civil Systems Engineering, Ajou University in Tashkent, Asalobod St. 113, Tashkent 100204, Uzbekistan}\address[addr4]{Webster University Tashkent, 13 Alisher Navoiy Avenue, Tashkent 100011, Uzbekistan}
\address[addr5]{National University of Uzbekistan, Tashkent 100174, Uzbekistan}\address[addr6]{Tashkent State Technical University, Tashkent 100095, Uzbekistan}

\begin{abstract}
The paper has explored analogue of gravitational synchrotron massive particle and Penrose process in MOdified Gravity (MOG) known as Scalar-Tensor-Vector-Gravity (STVG). Investigation of the gravitational field around Kerr-MOG black hole showed that it has strong gravitational field with large horizon and can rotate faster than Kerr black hole due to the effect of STVG. We have studied influence of STVG in circular motion of massive particle around Kerr-MOG black hole and discussed the Innermost Stable Circular Orbit (ISCO) of massive test particle. It is shown that STVG plays a crucial role in energy extraction from a rotating black hole, with an energy efficiency of more than $100\%$ according to the Penrose process. Furthermore, we have explored the gravitational synchrotron radiation analogue produced by a massive particle orbiting around a Kerr-MOG black hole. It has been shown that the intensity of gravitational radiation from binary systems of stellar black holes (SBH) and supermassive black holes (SMBH). 
\end{abstract}


\end{frontmatter}

\linenumbers


\section{Introduction}

Scalar-tensor-vector gravity (STVG) theory is a modification of Einstein's general relativity (GR) that incorporates an additional scalar and vector field. The theory was first proposed as a possible alternative to GR that could address some of the shortcomings of the theory, such as the need for dark matter and dark energy. In STVG, the scalar field is responsible for the expansion of the universe and is often associated with the Higgs boson. The vector field, on the other hand, is responsible for the gravitational interactions between particles and is often associated with the gauge bosons of the weak force. The theory is characterized by a set of field equations that describe the dynamics of the scalar and vector fields in the presence of matter and energy. These equations are more complex than the field equations of GR, but they can be solved numerically to make predictions about the behavior of gravitational systems. STVG, also known as MOdified Gravity (MOG), has been the subject of much research in recent years, and it is still an active area of investigation \cite{Moffat2006JCAP}. 

In later works MOG field equations were directly obtained from the least action principle and solutions of those equations were compared to cosmological and astrophysical oservations \cite{Moffat2006CQG}.
The theory was successfully applied for explanation of galaxy rotation curves \cite{Moffat2008ApJ}, cluster dynamics \cite{Brownstein2007MNRAS}, gravitational lensing \cite{Moffat2007IJMPD} and Solar system \cite{Moffat2007IJMPD}. There is a proposed explanation for the behavior of part-time pulsars by effects of MOG \citep{RayimbaevPRD2020}. MOG violates Birkhoff's theorem and some consequences of that were studied in \cite{Moffat2009MNRAS}. Results of the study might provide a new insight to the nature of inertia.

Further studies on gravitational lensing in point source spherically symmetric spacetimes were concluded with results that can be applied for the cases of continuous distributions of mass and provide a possible explanation for Abel 520 merging clusters \cite{Moffat2009MNRASa}. In \cite{Moffat2013Galaxies,Moffat2021EPJC} authors continue the studies and show that MOG can be used to model acoustic power spectrum of the cosmic microwave radiation and apply it to model the peak of the spectrum. Angular powr spectrum was studied in \cite{Moffat2021Universe}

Using the data of the HI Nearby Galaxy Survey catalogue of galaxies and the Ursa Major catalogue of galaxies a good fit to galaxy rotation curve was obtained and it was shown that mass-to-light ratio derived during fitting process is correlated with the colors of galaxies \cite{Moffat2013MNRAS,Moffat2014MNRAS}.

In \cite{Moffat2015EPJC} authors model the shadows of static spherically symmetric black holes in MOG, which is then done for rotating black holes in \cite{Moffat2015EPJCa} along with a construction of traversable wormhole solution. Black hole shadows in MOG were also investigated by \cite{Moffat2020PRD} Other optical properties of black holes, such as gravitational lensing and gravitational delay were studied by \cite{Rahvar2019MNRAS, Moffat2018Galaxies, Ovgun2019AP, Tuleganova2020GRG, Izmailov2019MNRAS}

Thermodynamical properties of black holes in MOG are analyzed in \cite{Mureika2016PLB}, results are similar to the Einstein-Maxwell solutions where the electric charge is replaced by a new parameter that depends on mass. Thermodynamics of regular solutions is also studied and horizonless limiting case is explored. Quasinormal modes were studied in MOG by \cite{Manfredi2017JPCS, Manfredi2018PLB,Manfredi2019JURP} and some fits to neutron star merger events were done by \cite{Green2018PLB}. Some studies on scalar fields in the vicinity of compact gravitational objects in MOG are done by \cite{Wondrak2018JCAP, Qiao2020EPJC, Moffat2021JCAP}. MOG was used by \cite{Duztas2020EPJC} to test the validity of the weak cosmic censorship conjuncture.

Fitting of gravitational wave data from LIGO-Virgo is performed in \cite{Moffat2016PLBa}, theory wa also applied to Abell 1689 galaxy cluster in \cite{Moffat2017EPJP} and newtonian and modified newtonian dynamics are shown to not fit the acceleration data were MOG and Navarro-Frenk-White dark matter models do. Some other studies on galaxy clusters and acceleration data include \cite{Moffat2019MNRAS, Moffat2019MNRAS, Green2019PDU}

There is a wide range of works on particle motion around compact gravitational objects in MOG focusing on different aspects, namely stable orbits \cite{Sharif2017EPJC,Lee2017EPJC, Perez2017PRD} and energy extraction \cite{Pradhan2019EPJC}. Particle motion approach was also utilized to study the motion of S2 star on its orbit around Sagittarius A* SMBH \cite{Monica2022Universe,Monica2022MNRAS,Turimov2022MNRAS,Monica2023MNRAS}, and constraints on parameters of the theory were discussed. Quasi-periodic oscillations in MOG were discussed by \cite{Kolos2020EPJC,Pradhan2019EPJC} and a new method for using QPO data to estimate alternative gravity theories parameters using QPOs was demonstrated using MOG by \cite{Rayimbaev2021Gal}. 

The Penrose process is a mechanism of energy extraction from a rotating black hole. The basic idea behind the Penrose process is that if a particle falls into a rotating black hole and splits into two fragments due to tidal forces in the region so-called ergosphere, some of its mass can be captured by the black hole, while the rest can be flung outwards with a higher energy than the original object had. This happens because the black hole's rotation imparts energy to the debris as it is flung outward, in a process known as frame dragging \cite{Penrose1971NPhS}. The Penrose process has important implications for astrophysics, as it can help explain the origin of high-energy particles observed in various astrophysical phenomena, such as active galactic nuclei and gamma-ray bursts. The maximum amount of energy gain for a single particle around the extreme Kerr black hole is $20.7\%$, while larger efficiencies are possible for a charged particle around magnetized and charged rotating black holes \cite{Dadhich2018MNRAS,Bhat1985JApA,Stuchlik2021Universe}. The Penrose process is not an efficient means of extracting energy from a black hole, but it demonstrates how black holes can generate enormous amounts of energy from their strong gravitational fields. It is also a key concept in the study of black holes and their properties in the framework of MOG theory. 

In the present paper, we are interested in investigating of dynamical motion of test particle around rotating black hole in STVG. We explicitly show the effect of the STVG in the ISCO position for massive particle. We also test STVG in gravitational analogue of the synchrotron radiation of massive particle and the Penrose process. The paper is organized as follows. In Sect.~\ref{Sec:1}, we study dynamical motion of massive particle around Kerr-MOG black hole. In Sect.~\ref{Sec:2}, we investigate the gravitational analogue of the synchrotron radiation from massive particle. In the next Sect.~\ref{Sec:3}, we test STVG with the Penrose process. Finally, in section~\ref{Sec:Conclusion} we summarize the main obtained results.

\section{Background spacetime and dynamics of particle}\label{Sec:1}

In the Boyer-Lindquist coordinates, the spacetime around rotating black hole in the STVG is described by the metric \cite{Moffat2015EPJC}: 
\begin{align}\label{metric}\nonumber
ds^2=&-\frac{\Delta}{\Sigma}\left(dt-a\sin^2\theta d\phi\right)^2+\frac{\Sigma}{\Delta}dr^2+\Sigma d\theta^2\\&+\frac{\sin^2\theta}{\Sigma}\left[(r^2+a^2)d\phi-adt\right]^2\ ,
\end{align}   
where $\Delta=r^2-2(1+\alpha)Mr+\alpha(1+\alpha)M^2+a^2$ and $\Sigma=r^2+a^2\cos^2\theta$. Here $M$ and $a$ are the total mass and specific angular momentum (spin) of the black hole while $\alpha$ is the parameter of the STVG theory. Notice that the associated vector potential, $\Phi_\alpha$, to the spacetime (\ref{metric}) which characterize the fifth interaction between test body with external vector field and this interaction is described by the Coulomb-like potential in the static spacetime. While in the rotating spacetime the vector potential can be extended as 
\begin{align}\label{pot}
\Phi_\mu=\frac{\sqrt{\alpha}Mr}{\Sigma}\left(-1,0,0,a\sin^2\theta\right)\ .  
\end{align} 
That means rotating black hole posses magneto dipole-type potential due to rotation.    

\begin{figure}	
\includegraphics[width=\columnwidth]{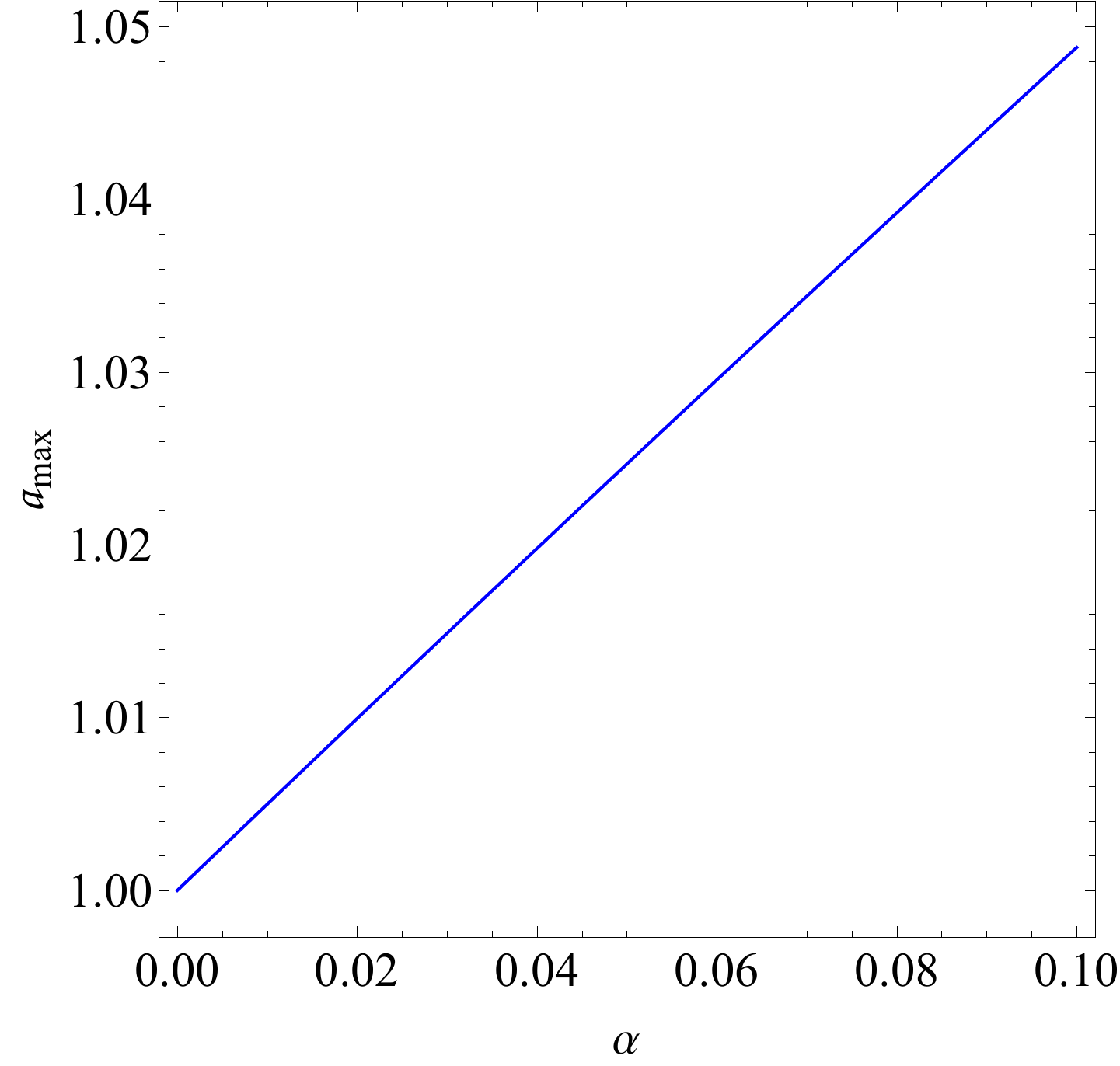}
\caption{Dependence of the maximal spin of the black hole from the STVG parameter $\alpha$ which is given by the equation $a_{\rm max}=\sqrt{1+\alpha}$.} \label{fig2}
\end{figure}

\begin{figure}	
\includegraphics[width=\columnwidth]{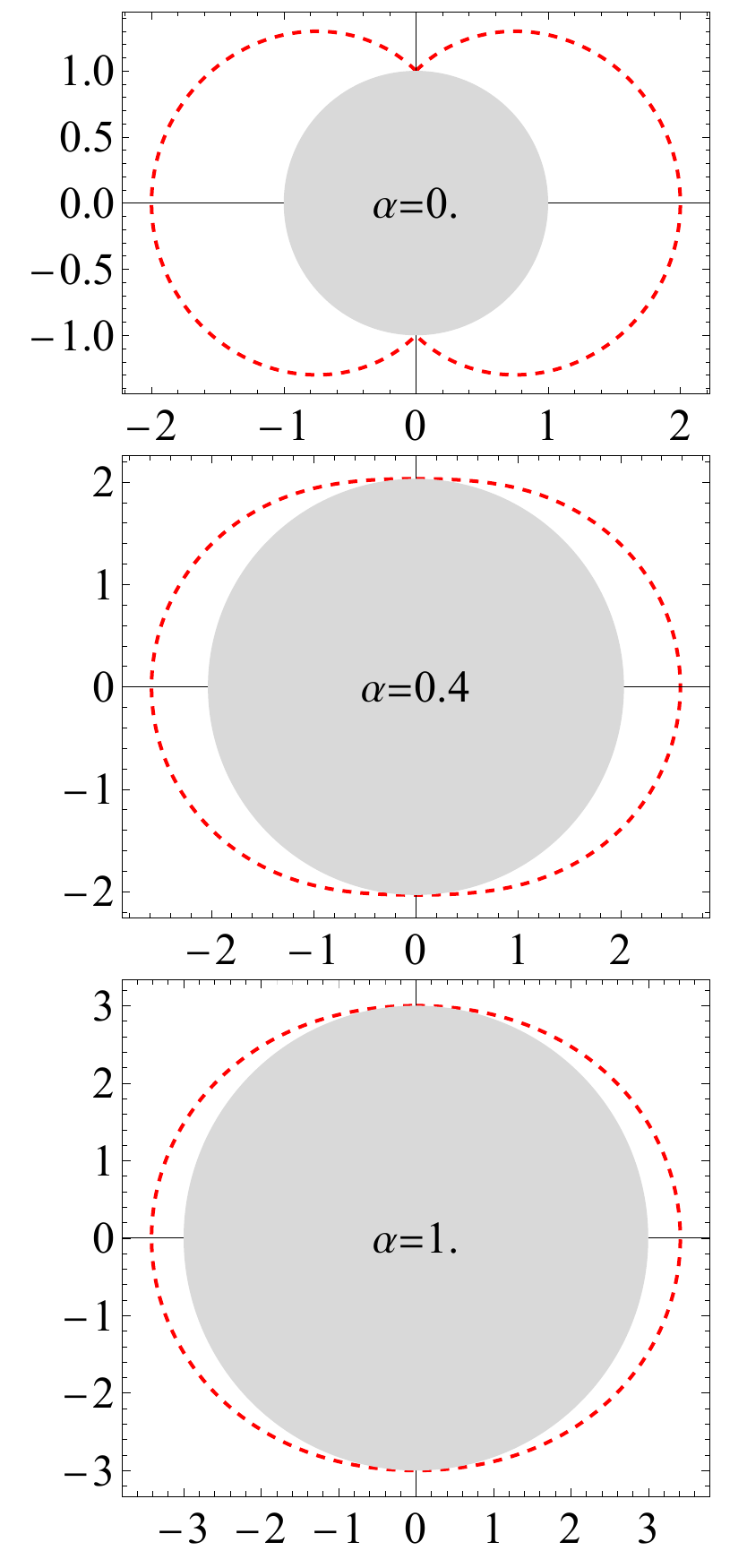}
\caption{The ergoshere of the black hole for the different values of the STVG parameter $\alpha$ for $a_*=1$.} \label{fig1}
\end{figure}

The horizon of the rotating black hole in STVG is located at $x_+=1+\alpha+\sqrt{1+\alpha-a_*^2}$, while the ergosphere of the black hole is determined as $x_e=1+\alpha+\sqrt{1+\alpha-a_*^2\cos^2\theta}$, where $x=r/M$ is a dimensionless radial coordinate and $a_*=a/M$ is a dimensionless spin parameter. The irreducible mass of the black hole in STVG is determined as $M_{\rm irr}=Mx_+/2$. According to the expression (\ref{pot}) the STVG parameter is always positive, i.e. $\alpha\geq 0$. From the existence of the black hole horizon one can find the limit the black hole's spin parameter as $|a_*|\leq\sqrt{1+\alpha}$ which means in the STVG theory the black hole can rotate faster than Kerr black hole. Dependence of the maximal spin of the black hole from STVG parameter $\alpha$ is illustrated in Fig. \ref{fig2}. It is known that the maximal spin of Kerr black hole is $a_{*\rm max}=1$ while in the STVG theory it equals to $a_{*\rm max}=\sqrt{1+\alpha}$. Figure \ref{fig1} draws ergosphere of Kerr-MOG black hole for different values of the STVG parameter $\alpha$. As one see that from this result that when the $\alpha$ parameter gets large the gravitational force dominates rotational force, therefore the region of the ergosphere gets smaller.

\begin{figure}	
\includegraphics[width=\columnwidth]{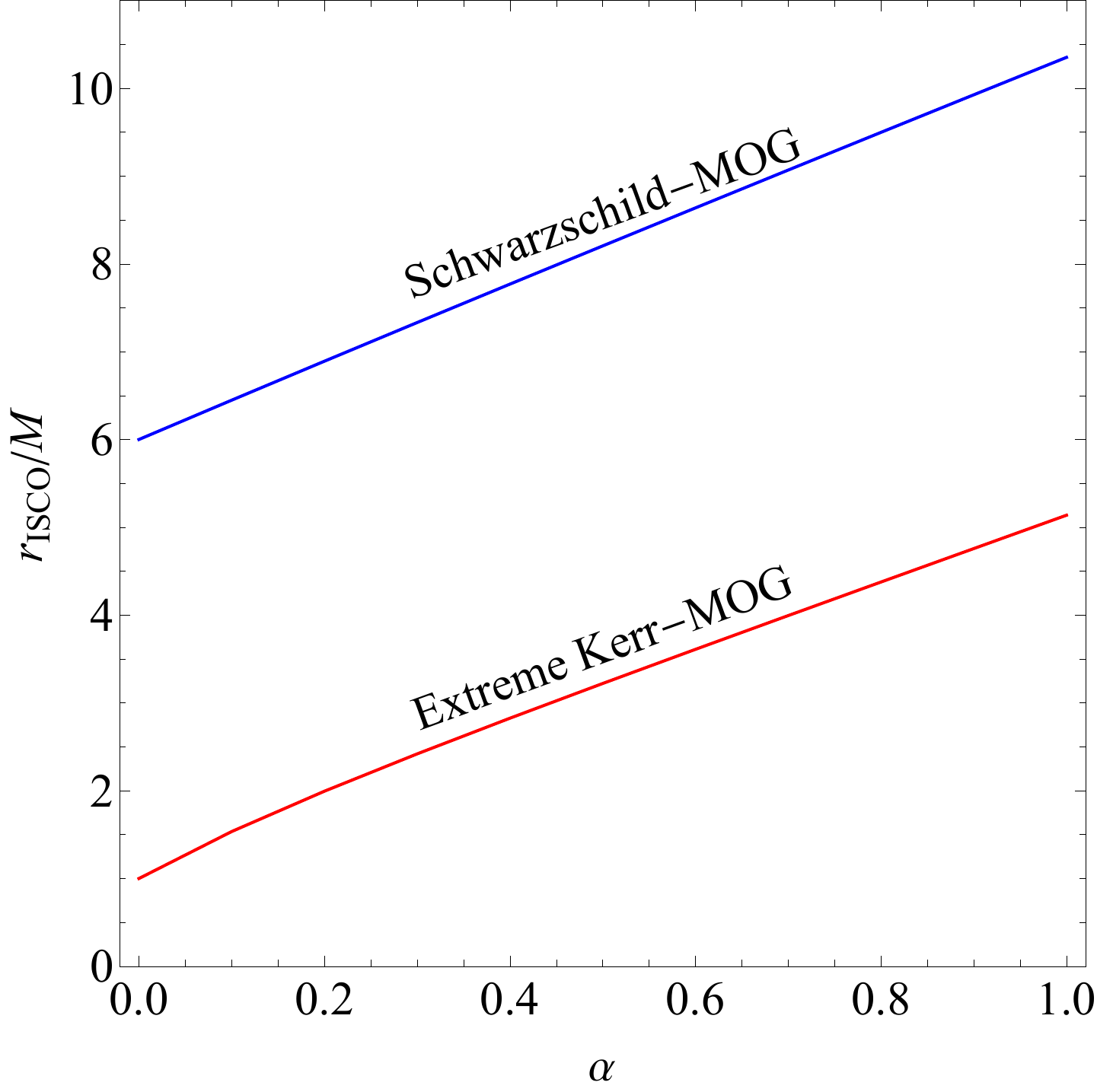}
\caption{Dependence of the ISCO position for massive particle from the STVG parameter $\alpha$ for static and extreme black holes.} \label{ISCO}
\end{figure}

Now we consider massive particle motion around black hole in the STVG theory. In this theory massive particles do not follow the geodesic line in unlike other gravity theories \cite{}. Therefore equation of motion for massive particle is governed by the following equation \cite{Moffat2015EPJCa}: 
\begin{align}\label{EOM}
{\ddot x}^\mu+\Gamma_{\nu\lambda}^\mu{\dot x}^\nu {\dot x}^\lambda=\frac{q}{m} B^\mu_{~\nu}{\dot x}^\nu \ ,\qquad {\dot x}^\mu=\frac{dx^\mu}{ds}\ ,
\end{align}
which can be derived from the following the Lagrangian  
\begin{align}
{\cal L}=\frac{1}{2}g_{\mu\nu}{\dot x}^\mu {\dot x}^\nu+\frac{q}{m} \Phi_\mu{\dot x}^\mu\ ,    
\end{align}
where $m$ is the mass of test particle, $q=\kappa m=\sqrt{\alpha}m$ is coupling constant of the interaction between the particle and the fifth force in STVG theory, ${\dot x}^\mu$ is the four-velocity of test particle normalized as $g_{\mu\nu}{\dot x}^\mu{\dot x}^\nu=-1$. An anti-symmetric tensor is defined as  $B_{\mu\nu}=\partial_\mu\Phi_\nu-\partial_\nu\Phi_\mu$.
The conserved quantities of motion, namely, the energy, $E$, and the angular momentum, $L$, of test particle are satisfied the following relations: 
\begin{align}
&g_{tt}{\dot t}+g_{t\phi}{\dot\phi}=-\frac{E+q\Phi_t}{m}\ ,\quad
&g_{\phi \phi}{\dot \phi}+g_{t\phi}{\dot t}=\frac{L-q\Phi_\phi}{m}\ ,
\end{align}
%
%
Using the normalization of the four-velocity and taking into account above expressions,  one can have the following equation:
\begin{align}\nonumber
&\left(E+q\Phi_t\right)^2-2\omega\left(E+q\Phi_t\right)\left(L-q\Phi_\phi\right)\\&+\frac{g_{tt}}{g_{\phi\phi}}\left(L-q\Phi_\phi\right)^2+m^2\left(g_{rr}\dot r^2+g_{\theta\theta}\dot\theta^2+1\right)\psi=0 \ .
\end{align}
where $\omega=-g_{t\phi}/g_{\phi\phi}>0$ and $\psi=g_{tt}-g_{t\phi}^2/g_{\phi\phi}<0$. Considering motion of massive particle in circular orbit with the following conditions $\dot r=0$ and $\dot\theta=0$, one can find that in STVG, massive particle motion is bounded by the following effective potential \cite{Dadhich2018MNRAS}:
\begin{align}
V=-q\Phi_t+\omega\left(L-q\Phi_\phi\right)+\sqrt{-\psi\left[m^2+\frac{\left(L-q\Phi_\phi\right)^2}{g_{\phi\phi}}\right]}\ ,
\end{align}
satisfying the following condition $E=V$. It is well-known that the effective potential is a useful tool for understanding the motion of test particles around black holes, and the position of the innermost stable circular orbit (ISCO) can be determined by finding the minimum value of the effective potential. The ISCO position of massive particle around a black hole is the smallest possible orbit where a test particle can maintain a stable circular orbit without being drawn into the black hole or flung away into space. The ISCO is a significant location for astrophysical observations, as it can affect the emission of radiation and matter falling onto the black hole. The study of the ISCO can also provide insights into the properties of black holes and tests of general relativity including the alternative theories of gravity. The ISCO position depends on the mass and spin of the black hole, as well as the STVG parameter in the present research work. Next thing what we discuss is finding the radius of the ISCO of massive particle around the black hole in STVG and to test how does it depends on the $\alpha$ parameter. For simplicity, we consider motion of massive particle in the equatorial plane that means the effective potential depends on the radial coordinate only, i.e. $V=V(r)$. Using the standard method the ISCO position can be found from the following conditions: $V(r)=E$, $V'(r)=0$ and $V''(r)\leq 0$. Obtaining the analytical expression for the ISCO position, therefore careful numerical analyses showed that the ISCO position gets larger due the effect of STVG. Figure \ref{ISCO} shows dependence of the ISCO position of massive particle orbiting around the Schwarzschild-MOG  and extreme Kerr-MOG black holes from the $\alpha$ parameter. Notice that the ISCO position is located in the region between two curves in Fig. \ref{ISCO} for arbitrary values of the spin parameter with range of $0<a_*<\sqrt{1+\alpha}$.

\section{Gravitational synchrotron radiation}\label{Sec:2}

It is well established that accelerating charged particles emit synchrotron radiation, which is a form of electromagnetic radiation that is emitted when a charged particle is accelerated in a curved path in the presence of a magnetic field. The intensity of synchrotron radiation emitted by the charged particle is proportional to the square of the acceleration $I\sim w_\mu w^\mu$ and the frequency of the emitted radiation depends on the strength of the magnetic field. It is interesting to note that the synchrotron radiation emitted by relativistic charged particles around magnetized and charged black holes has been the subject of recent research. This is because black holes are known to have strong gravitational and electromagnetic fields that can affect the motion of charged particles and alter the characteristics of the emitted radiation. In the Refs. \cite{Tursunov2018AN,Tursunov2018ApJ,Kolos2021PRD,Turimov2022EPJP}, it is provided insights into the synchrotron radiation spectrum of relativistic charged particles around black holes, and their results have important implications for our understanding of the astrophysical phenomena associated with black holes.

\begin{figure}	
\includegraphics[width=\columnwidth]{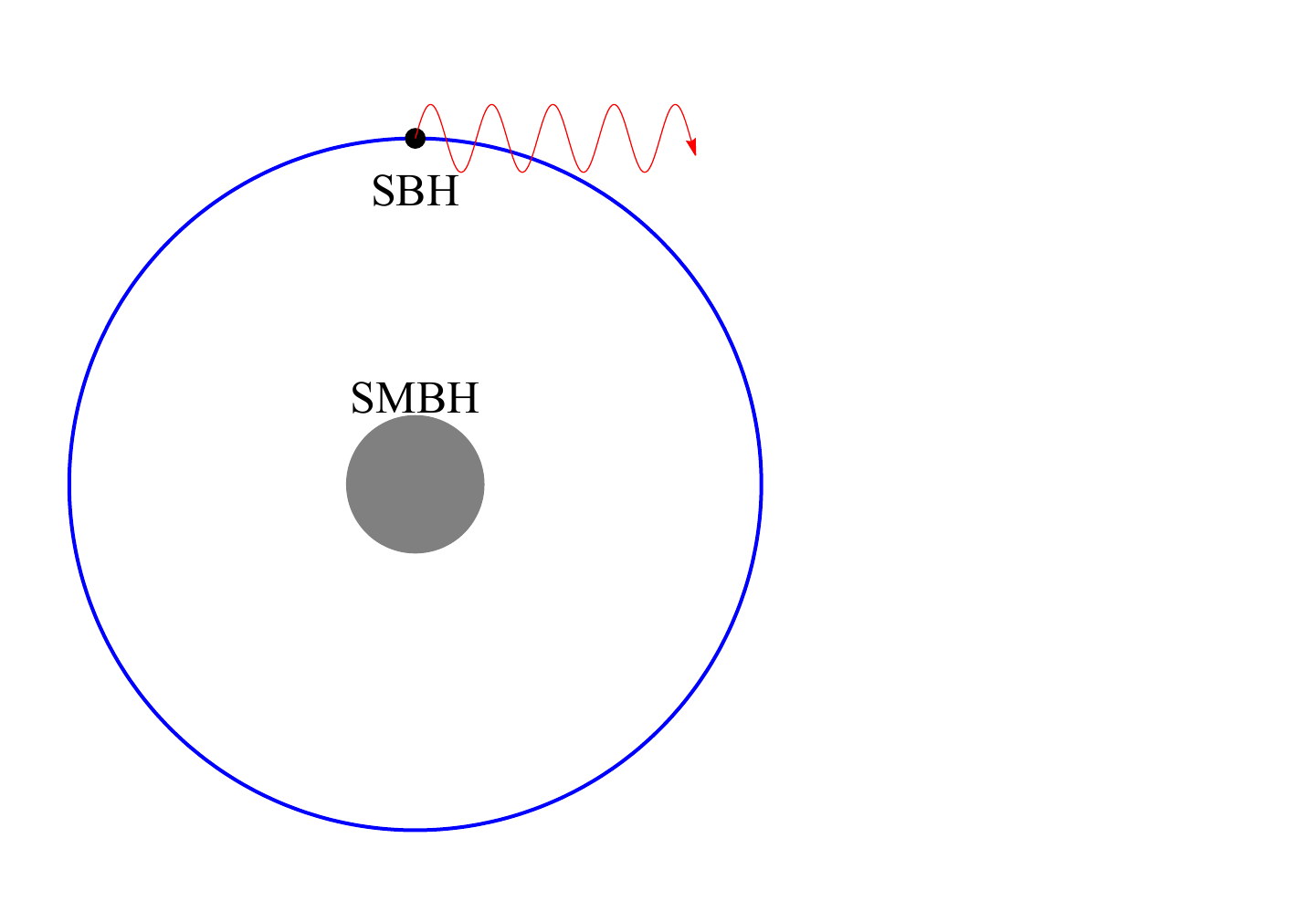}
\caption{The schematical picture of radiated SBH around SMBH in STVG.} \label{fig3}
\end{figure}

Here we will discuss gravitational analogue of the synchrotron radiation from massive particle orbiting around black hole in STVG. As we mentioned before that in this theory, massive particle does not follow geodesics line due to the fifth force and from equation of motion in (\ref{EOM}), the four-acceleration of massive particle is given as
\begin{align}\label{acceleration}
w^\mu=\frac{q}{m}B^\mu_{~\nu}{\dot x}^\nu\ , \qquad w_\mu{\dot x}^\nu=0\ ,  
\end{align}
which is always perpendicular to the four-velocity of particle. Notice that equation \eqref{acceleration} is gravitational analogue of the Lorentz equation, and charge of particle is replaced by the coupling constant $q$, while anti-symmetric tensor $B_{\mu\nu}$ stands instead of the Faraday tensor in Lorentz equation.  

Our main assumption is that if massive particle orbits with non-zero value of the four-acceleration in the presence of the external force then it should emits radiation.  We call this radiation as {\it gravitational synchrotron radiation}. Then the intensity of accelerating massive particle in STVG is determined as 
\begin{align}
I&=-\frac{2q^2}{3}w_\mu w^\mu=\frac{2q^4}{3m^2}B_{\mu\lambda}B^{\mu\nu}{\dot x}^\lambda {\dot x}_\nu\ . \end{align}

Considering circular motion of a massive particle with four-velocity of ${\dot x}^\mu={\dot t}(1,v,0,\Omega)$, where $v=dr/dt$ and $\Omega=d\phi/dt$ are radial and angular velocities of massive particle. Hereafter, performing simple algebra manipulations one can obtain the expression for the acceleration the following the form:
\begin{align}
&w_\mu=\frac{q}{m}{\dot t}(vB_{tr}, B_{rt}+\Omega B_{r\phi}, B_{\theta t}+\Omega B_{\theta\phi}, vB_{\phi r})\ ,
\end{align}
where non-zero components of the anti-symmetric tensor can be found as
\begin{align}
&B_{rt}=\frac{\sqrt{\alpha}M}{\Sigma^2}(r^2-a^2\cos^2\theta)\ ,\qquad B_{r\phi}=-B_{rt}a\sin^2\theta\ ,
\\
&B_{\theta t}=-\frac{\sqrt{\alpha}Mra^2\sin 2\theta}{\Sigma^2}\ ,\qquad B_{\theta\phi}=-B_{\theta t}\frac{r^2+a^2}{a}\ ,
\end{align}
and $\dot t$ can be found using normalization of the four-velocity ${\dot t}^{-1}=\sqrt{-g_{tt}-2\Omega g_{t\phi}-\Omega^2g_{\phi\phi}-v^2g_{rr}}$.

\begin{figure}	
\includegraphics[width=\columnwidth]{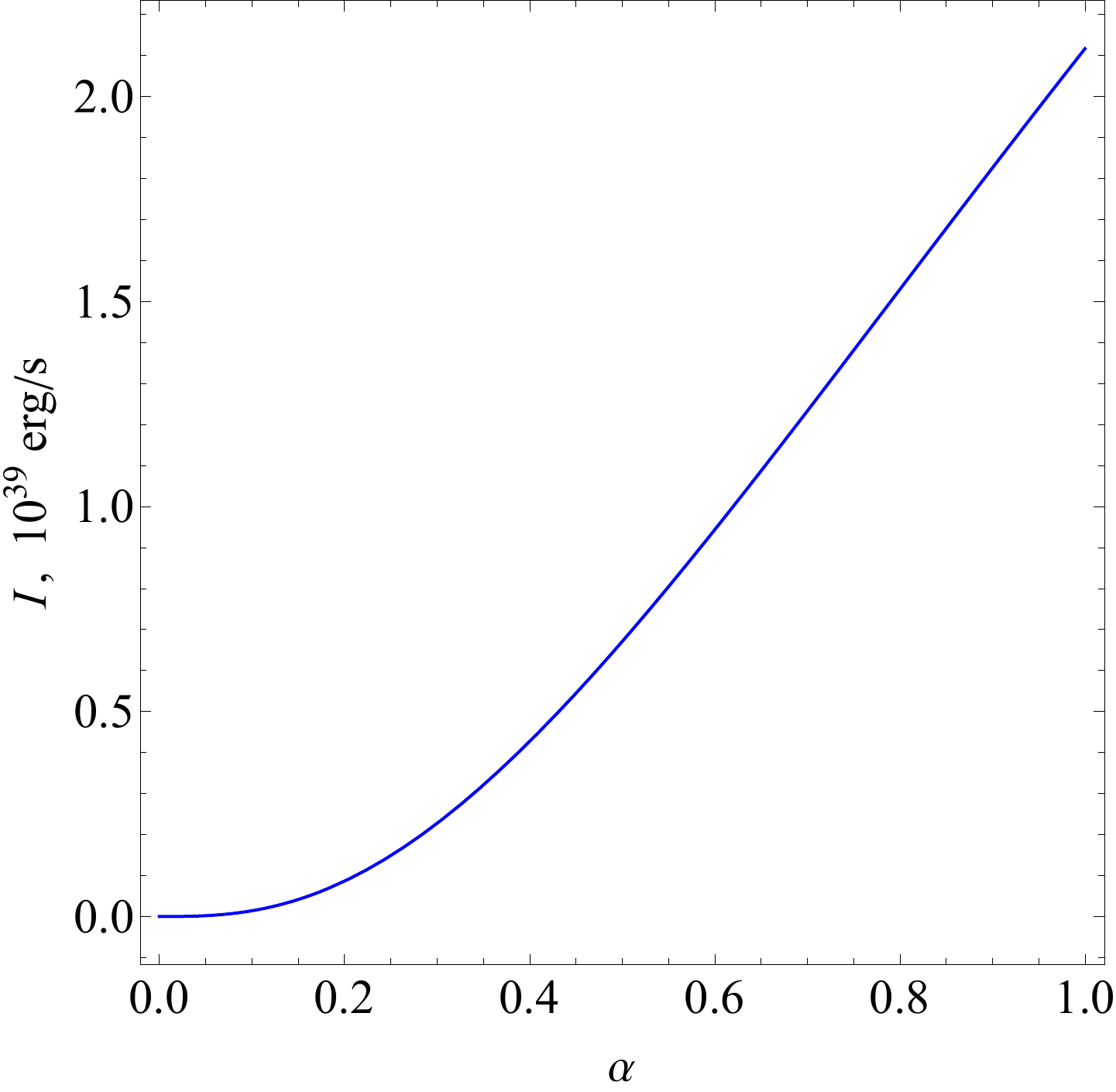}
\caption{Dependence of the intensity of a radiating SBH at the ISCO position from the $\alpha$ parameter.} \label{Intensity}
\end{figure}

The expression for the intensity of the radiating massive particle orbiting around Kerr-MOG black hole in equatorial plane reads
\begin{align}
I&=\frac{2\alpha^3G^3m^2M^2}{3c^3r^4}{\dot t}^2\left[\frac{\Delta}{r^2}\left(1-\Omega a\right)^2-\frac{r^2}{\Delta}\frac{v^2}{c^2}\right]\ .  
\end{align}
As one can see from above equation that massive particle emits gravitational radiation even in arbitrary stable orbit (i.e. $v=0$) and in the case of a non-rotating black hole ${\dot t}^2$ can be canceled with expression in bracket. In Fig. \ref{fig3} a schematically picture of gravitational radiating SBH in the vicinity of a SMBH is illustrated. As a result one can estimate the intensity of gravitational radiation from the stellar black hole (SBH) orbiting around supermassive black hole (SMBH) as
\begin{align}\label{estimation}
I\sim 2.43\times 10^{39}\alpha^3\left(\frac{m}{10M_\odot}\right)^2\left(\frac{M}{10^9 M_\odot}\right)^{-2}\left(\frac{10GM}{c^2r}\right)^4{\rm erg/s}\ ,  
\end{align}
which depends on the radial coordinate. It is also interesting to calculate the intensity at the ISCO position and determine how it depends on the $\alpha$ parameter. Figure \ref{Intensity} shows the dependence of the intensity on the $\alpha$ parameter. As one can see from the figure, in the absence of the STVG parameter, the intensity becomes zero because, in this case, the massive particle follows a geodesic line near the black hole and does not emit gravitational radiation. However, in the presence of the STVG (i.e., $\alpha \neq 0$), the intensity of the radiating particle becomes non-zero and increases with an increase in the $\alpha$ parameter.

\section{Penrose process}\label{Sec:3}

The Penrose process involves taking advantage of the strong gravitational field and rotation of a black hole to extract energy from it. In this process, a particle or a photon falls into the black hole's ergosphere, which is the region just outside the event horizon where spacetime is dragged along with the black hole's rotation. According to the Penrose process the particle falling onto the black hole separates into two parts in the ergosphere, with one part falling into the black hole and the other part escaping outwards. The part that escapes carries away some of the black hole's rotational energy, resulting in a net loss of energy for the black hole \cite{Penrose1971NPhS}. This extracted energy can then be used to power various astrophysical phenomena. The Penrose process is an important concept in theoretical astrophysics and is used to explain some of the most energetic phenomena in the universe. 

In the present paper, we are aimed to test STVG theory with the Penrose process. The energy efficiency of the particle is depends of the mass, spin parameter and the STVG parameter. Assume that massive particle (1) of parameters $(m_1,E_1,L_1,\dot{r}_1,\dot\theta_1,\dot\phi_1)$ falls onto black hole from infinity and decays into two fragments (2) and (3) with parameters of $(m_2,E_2,L_2,\dot{r}_2,\dot\theta_2,\dot\phi_2)$ and $(m_3,E_3,L_3,\dot{r}_3,\dot\theta_3,\dot\phi_3)$ in the ergosphere of the rotating black hole, where $m_i$, $E_i$, $L_i$, $\dot r_i$ and $\dot\theta_i$ are, respectively, the mass, energy, angular momentum, radial, vertical and azimuthal velocities. The conservation laws for this process can be written as
\begin{align}&\label{con}
E_1=E_2+E_3\ ,\quad L_1=L_2+L_3\ , \quad m_1\geq m_2+m_3 \ ,\\ &m_1\dot{r}_1=m_2\dot{r}_2+m_3 \dot{r}_3\ ,\quad 0=m_2\dot\theta_2+m_3\dot\theta_3\ ,\\\label{phidot} &m_1\dot\phi_1=m_2\dot\phi_2+m_3\dot\phi_3\ .
\end{align}

Again we consider circular motion of a massive particle with four-velocity of ${\dot x}^\mu={\dot t}(1,v,0,\Omega)$. Using the normalization of the four-velocity, the angular velocity of particle (1) is determined as \cite{Tursunov2020ApJ}
\begin{align}
\Omega_1=\frac{-(u^2+g_{tt})g_{t\phi}+u\sqrt{(-\psi)(u^2+g_{tt})g_{\phi\phi}}}{u^2g_{\phi\phi}+g_{t\phi}^2} \ ,  
\end{align}
where $u=(E_1+q_1\Phi_t)/m_1$, while the  angular velocity of the splitted fragments denote $\Omega_2=\Omega_+$ and $\Omega_3=\Omega_-$, where $\Omega_\pm$ are defined as 
\begin{align}
\Omega_\pm=-\frac{g_{t\phi}}{g_{\phi\phi}}\pm\sqrt{\left(\frac{g_{t\phi}}{g_{\phi\phi}}\right)^2-\frac{g_{tt}}{g_{\phi\phi}}}\ .   
\end{align}
\begin{figure}	
\includegraphics[width=\columnwidth]{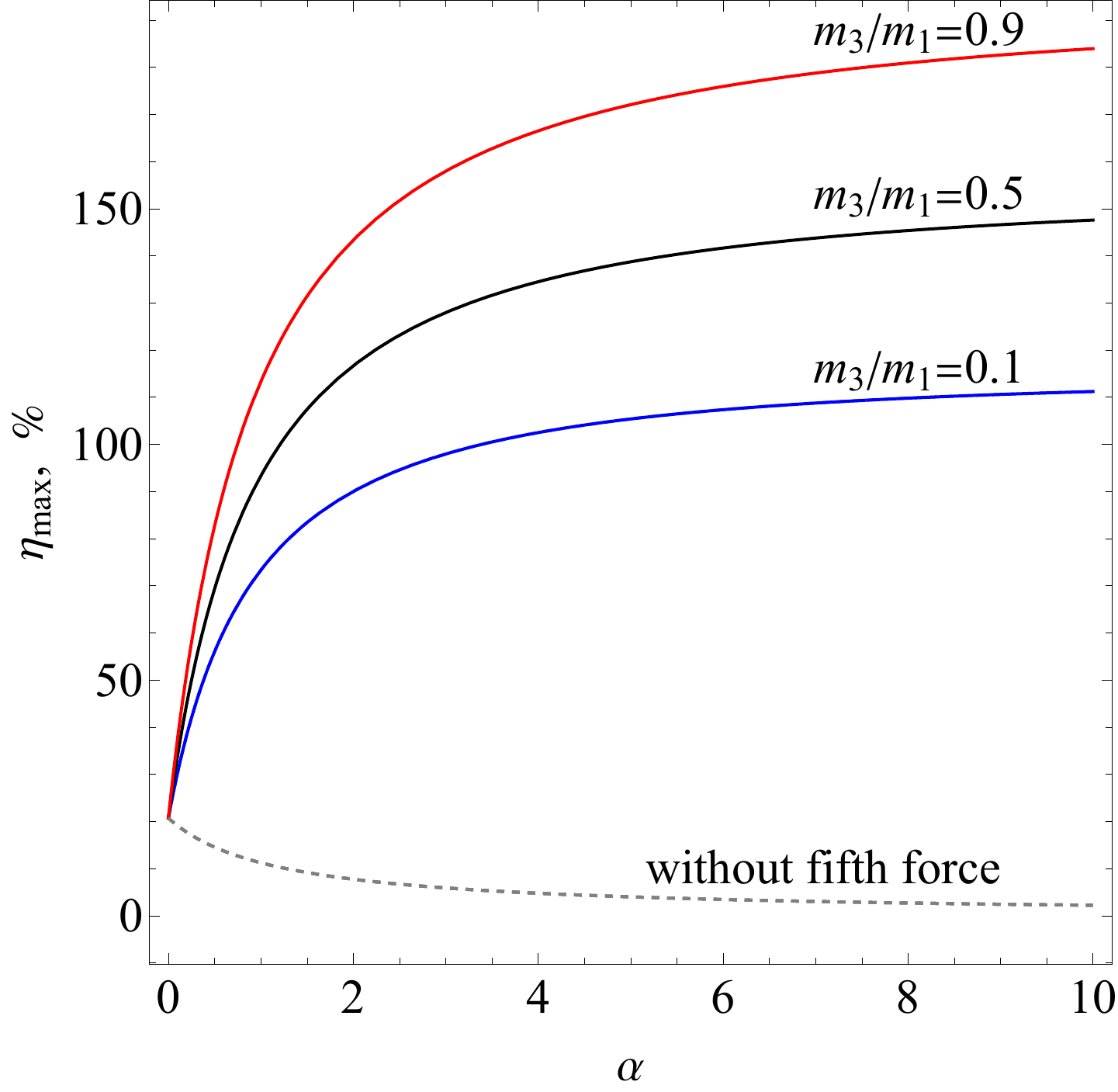}
\caption{Dependence of the maximal energy efficiency of the black hole from the STVG parameter $\alpha$.} \label{fig4}
\end{figure}

From the inequality for particles mass in \eqref{con}, relation between coupling constants can be obtained as $q_1\geq q_2+q_3$. Using equation \eqref{phidot} and after simple algebraic manipulations, the relation between energies of in falling and escaping particles can be found as 
\begin{align}\label{relation}
\left(E_3+q_3\Phi_t\right)\frac{\Omega_3-\Omega_2}{g_{tt}+\Omega_3 g_{t\phi}}\geq \left(E_1+q_1\Phi_t\right)\frac{\Omega_1-\Omega_2}{g_{tt}+\Omega_1 g_{t\phi}}\ .   
\end{align}

The energy efficiency is determined as $\eta=E_3/E_1-1$, while in STVG theory is
\begin{align}\label{EE}\nonumber
\eta&\geq\left(1+\frac{q_1\Phi_t}{E_1}\right)\frac{\Omega_1-\Omega_2}{\Omega_3-\Omega_2}\frac{g_{tt}+\Omega_3g_{t\phi}}{g_{tt}+\Omega_1 g_{t\phi}}-\frac{q_3\Phi_t}{E_1}-1\\&=\frac{1}{2}\left(\sqrt{1+\frac{g_{tt}}{u^2}}-1\right)\left(1+\frac{q_1}{E_1}\Phi_t\right)-\frac{q_1+q_3}{E_1}\Phi_t\ .  
\end{align}
To make qualitative analyses of the energy efficiency in STVG, one can estimate the energy of falling particle as $E_1\simeq m_1$ and the expression \eqref{EE} for the extreme rotating black hole (i.e. $x_+=1+\alpha$) reads  
\begin{align}&
\eta_{\rm max}\geq\frac{1}{2(1+\alpha)}\left[\sqrt{2+\alpha}-1+2\alpha\left(1+\frac{m_3}{m_1}\right)\right]\ .  
\end{align}
For large value of the STVG parameter, i.e. $\alpha\to\infty$, the energy efficiency is $\eta_{\rm max}\geq 1+m_3/m_1$, which is greater than $100\%$ because of mass ratio of escaping and falling particles, while in the case of the extreme Kerr black black hole, $\eta_{\rm max}=(\sqrt{2}-1)/2\simeq 0.207$, which is around $\sim 21\%$. Figure \ref{fig4} shows that maximal efficiency of energy extraction from the Kerr-MOG black hole reaches up to $\sim 200\%$ for particular value of mass ratio of escaping and falling which is quite huge. However, in the research paper \cite{Dadhich2018MNRAS}, it has been demonstrated that a magnetized Kerr black hole is considered one of the most promising candidates for accelerating high-energy particles. The study reports that the efficiency of energy extraction from a magnetized Kerr black hole can exceed $100\%$ due to the induced electrostatic potential resulting from rotation and magnetic field effects. Likewise, in the framework of STVG theory, the external fifth-potential plays a crucial role in generating high-energy particles and potentially surpassing the $100\%$ efficiency limit in energy extraction from a Kerr-MOG black hole.

\begin{figure}	
\includegraphics[width=\columnwidth]{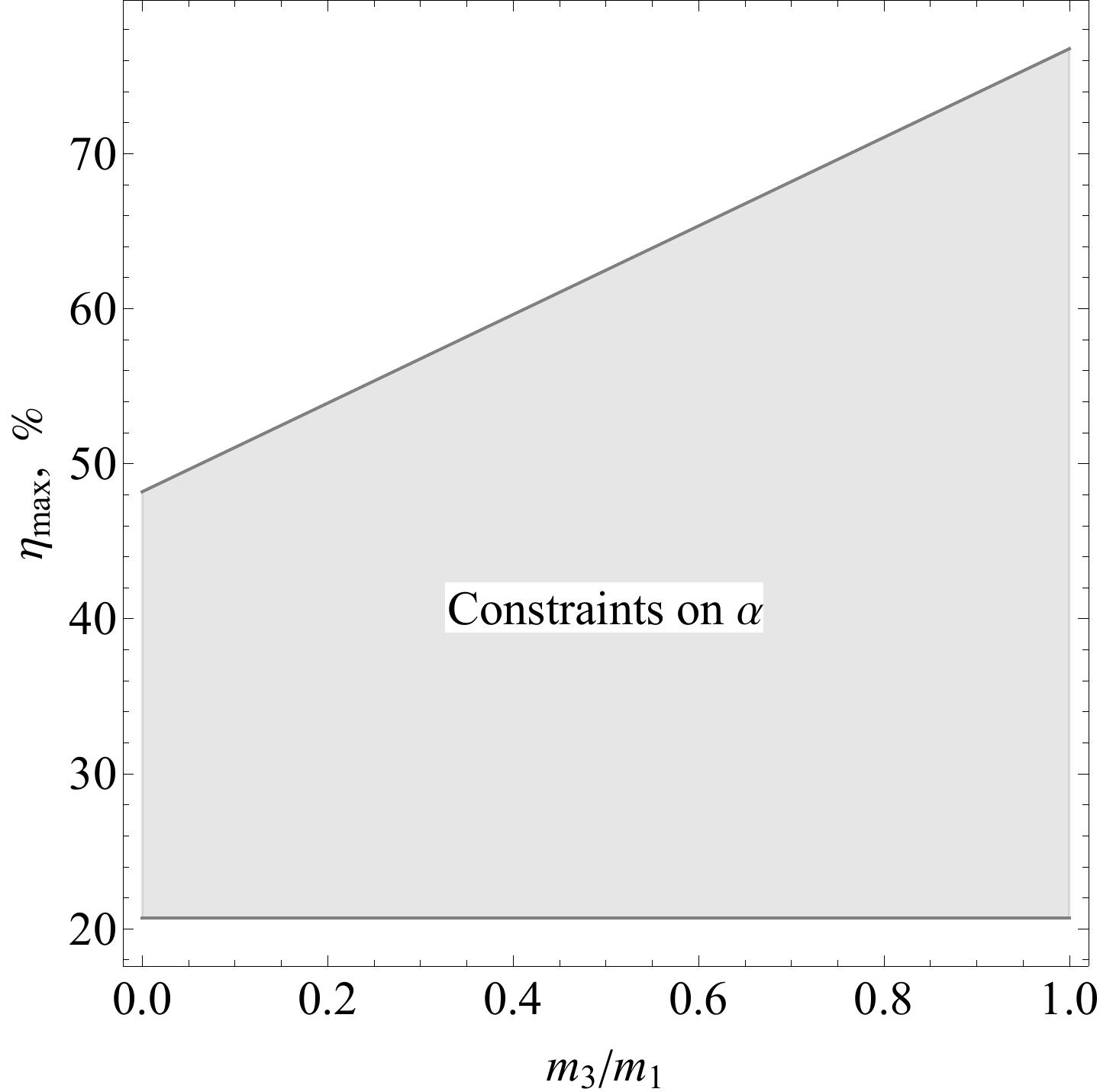}
\caption{Dependence of the maximal energy efficiency of the black hole from mass ratio of escaping and falling particles at $\alpha\lesssim 0.41$.} \label{fig5}
\end{figure}

We see that under some circumstances the efficiency of energy extraction via Penrose process can be very large, exceeding $>100\%$. This might open the possibilities to obtain additional constraints on MOG parameter $\alpha$ which might be a good topic for further research. In the paper \cite{RayimbaevPRD2020}, the authors have suggested that MOG may have implications for the radio emission properties of pulsars. The upper limit for the parameter $\alpha$ in MOG using observational data from specific millisecond pulsars has been estimated. For the millisecond pulsar J2145-0750, the estimated upper limit for $\alpha$ is $\sim 1.6011$. Similarly, for the millisecond pulsars J0024-7204D and J0024-7204H, the estimated upper limits for $\alpha$ are $\sim 3.06528$ and $\sim 0.9747$, respectively. However these values of MOG parameter are quite huge for the black hole. In Ref. \cite{Monica2022MNRAS}, the constraint on MOG parameter $\alpha$  has been obtained considering the motion of the S2-star around the supermassive black hole at the center of the Milky Way described Schwarzschild-MOG spacetime and by comparing the predicted orbital motion of the S2-star based on the with the available astrometric data, including measurements of positions, radial velocities, and orbital precession consisting of $145$ position measurements, $44$ radial velocity measurements. Using a Monte Carlo Markov Chain (MCMC) algorithm, it has been found that the  constrained to be $\alpha\lesssim 0.41$ at the $99.7\%$ confidence level. Using this upper value for MOG parameter dependence of the maximum efficiency of energy extraction via Penrose process from mass ratio of escaping and falling particle is illustrated in Fig. \ref{fig5} and in this case, the maximum energy efficiency reaches up to $\sim 78\%$ at $\alpha\lesssim 0.41$.

One can also discuss the very similar mechanism around rotating black hole in the STVG which is known as the Banados-Silk-West (BSW) process which inverse process to the Penrose one. This process is a theoretical mechanism for the creation of a black hole in the collision of two high-energy particles \cite{BSW2009PRL}. The BSW process is significant because it provides a new way to study the properties of black holes and the nature of gravity. It also has implications for the study of high-energy physics and the behavior of particles at extremely high energies.  

The BSW process involves the collision of two high-energy massive particles. In this process, the size of the resulting black hole depends on the energy of the colliding particles, which can be related to their mass and velocity through the principles of special relativity. The exact equations and calculations involved in the BSW process are complex and beyond the scope of a simple answer, but they involve the principles of general relativity, quantum mechanics, and particle physics. According to this process the energy of centre mass of two collision particle is given by $E_{\rm cm}^2=2m_0^2(1-g_{\mu\nu}u_1^\mu u_2^\nu)$, where $m_0$ is the mass of the particles at the infinity. In the STVG, it can be represented as
\begin{align}\label{Ecm}
&\frac{E_{\rm cm}^2}{2m_0^2}=1-\frac{a^2}{r^2}+\frac{T^2-(2+\alpha)aMr(l_1+l_2)+(a^2-\Delta)l_1l_2}{\Delta r^2}\\\nonumber&-\sqrt{\left[\frac{(T-al_1)^2}{r^2\Delta}-\frac{\left(l_1-a\right)^2}{r^2}-1\right]\left[\frac{(T-al_2)^2}{r^2\Delta}-\frac{\left(l_2-a\right)^2}{r^2}-1\right]}\ ,
\end{align}
where $T=r^2+a^2-\alpha Mr$. This expression divergences at the horizon because of $\Delta=0$, however, our analyses showed that numerator of the expression also divergences. Therefore near the horizon the expression \eqref{Ecm} reduces to
\begin{align}\nonumber
\frac{E_{\rm cm}^2}{2m_0^2}&=\frac{l_1-2\sqrt{1+\alpha}}{l_2-2\sqrt{1+\alpha}}+\frac{l_2-2\sqrt{1+\alpha}}{l_1-2\sqrt{1+\alpha}}\\&-\frac{\alpha(l_2-l_1)^2}{2(1+\alpha)(l_1-2\sqrt{1+\alpha})(l_2-2\sqrt{1+\alpha})}\ ,
\end{align}
for the equal values of the angular momentum $l_1=l_2$, the energy of the centre mass of the system will be $E_{\rm cm}=2m_0$. The radial dependence of the energy of the centre mass showed that it tends to infinity, that means the Kerr-MOG black hole might be a source of the high-energy particles.  

\section{Conclusions}\label{Sec:Conclusion}

It is well-established that the motion of massive particles around black holes in STVG does not follow a geodesic line. Building upon this fact, we have conducted a study of the circular motion of massive test particles in the vicinity of the Kerr-MOG black hole, which is one of the black hole solutions in STVG. To simplify our analysis, we have focused on circular motion in the equatorial plane and have derived the effective potential governing the motion of the massive particle in orbit around the Kerr-MOG black hole. Through numerical calculations, we have discovered that the position of the ISCO (innermost stable circular orbit) for the massive particle around the black hole increases due to an external force arising from STVG. We have also graphically demonstrated the dependence of the ISCO position on the STVG parameter $\alpha$. Overall, our research sheds light on the behavior of massive particles orbiting around black holes in the context of STVG theory, providing new insights into the effects of external forces on the motion of these particles.      

As we mentioned before, massive particles occurs four-acceleration due to the fifth force in STVG, and the equation of motion resembles the Lorentz equation for charged particles. Based on this fact, we have discussed the gravitational analogue of synchrotron radiation from a massive particle orbiting around a Kerr-MOG black hole due to the fifth force in STVG. Additionally, we have studied the gravitational radiation emitted by a massive particle falling radially onto a Kerr-MOG black hole with angular velocity. We are arguing that massive particle, such as black hole, orbiting around the SMBH, can emit intense radiation and generate gravitational waves that can be observed. This effect is known as the "gravitational radiation reaction," and it is caused by the particle's motion in the strong gravitational field of the SMBH. The intensity of the radiation emitted by the orbiting particle depends on various factors, such as its mass, velocity, and distance from the black hole. Based on our analysis, for a typical SMBH with a mass of $10^9 M_\odot$ and SBH of mass $10M_\odot$ orbiting around it, the intensity of the radiation has been estimated to be $I\sim 2.43\times 10^{39}$. This is a very large value and suggests that such systems could be strong sources of gravitational waves. Indeed, the observation of gravitational waves from binary black hole mergers by the LIGO and Virgo collaborations has opened up a new window into the study of black holes and their environments in the STVG. The detection of gravitational waves from other types of black hole systems, such as those with orbiting massive particles, would provide further insights into the physics of black holes and their interactions with their surroundings. 

It has been tested STVG with the Penrose process and analyzed its impact on the energy extraction from a Kerr-MOG black hole. The Penrose process is a phenomenon where energy can be extracted from a rotating black hole. It is well-known that the energy efficiency of such process for extreme Kerr black hole is about $\sim 21\%$, while in the presence of the STVG it is greater than $100\%$. It is also discussed the BSW (Banados-Silk-West) process around the Kerr-MOG black hole and suggested that the energy of the center of mass of two colliding particles increases significantly. This is also an interesting finding, as it suggests that STVG could potentially have implications for high-energy collisions. This work presents valuable opportunities for future research works in this area.

\section*{Acknowledgements}
This research was supported by the Grants F-FA-2021-510 and MRB-2021-527 from the Uzbekistan Ministry for Innovative Development.


\bibliographystyle{model1-num-names}
\bibliography{example} 

\begin{thebibliography}{54}
\expandafter\ifx\csname natexlab\endcsname\relax\def\natexlab#1{#1}\fi
\providecommand{\bibinfo}[2]{#2}
\ifx\xfnm\relax \def\xfnm[#1]{\unskip,\space#1}\fi
\bibitem[{{Moffat}(2006{\natexlab{a}})}]{Moffat2006JCAP}
\bibinfo{author}{J.~W. {Moffat}},
\newblock \bibinfo{title}{{Scalar tensor vector gravity theory}},
\newblock \bibinfo{journal}{JCAP} \bibinfo{volume}{2006}
  (\bibinfo{year}{2006}{\natexlab{a}}) \bibinfo{pages}{004}.
\bibitem[{{Moffat}(2006{\natexlab{b}})}]{Moffat2006CQG}
\bibinfo{author}{J.~W. {Moffat}},
\newblock \bibinfo{title}{{Time delay predictions in a modified gravity
  theory}},
\newblock \bibinfo{journal}{Classical and Quantum Gravity} \bibinfo{volume}{23}
  (\bibinfo{year}{2006}{\natexlab{b}}) \bibinfo{pages}{6767--6771}.
\bibitem[{{Moffat} and {Toth}(2008)}]{Moffat2008ApJ}
\bibinfo{author}{J.~W. {Moffat}}, \bibinfo{author}{V.~T. {Toth}},
\newblock \bibinfo{title}{{Testing Modified Gravity with Globular Cluster
  Velocity Dispersions}},
\newblock \bibinfo{journal}{ApJ} \bibinfo{volume}{680} (\bibinfo{year}{2008})
  \bibinfo{pages}{1158--1161}.
\bibitem[{{Brownstein} and {Moffat}(2007)}]{Brownstein2007MNRAS}
\bibinfo{author}{J.~R. {Brownstein}}, \bibinfo{author}{J.~W. {Moffat}},
\newblock \bibinfo{title}{{The Bullet Cluster 1E0657-558 evidence shows
  modified gravity in the absence of dark matter}},
\newblock \bibinfo{journal}{MNRAS} \bibinfo{volume}{382} (\bibinfo{year}{2007})
  \bibinfo{pages}{29--47}.
\bibitem[{{Moffat}(2007)}]{Moffat2007IJMPD}
\bibinfo{author}{J.~W. {Moffat}},
\newblock \bibinfo{title}{{a Modified Gravity and its Consequences for the
  Solar System, Astrophysics and Cosmology}},
\newblock \bibinfo{journal}{International Journal of Modern Physics D}
  \bibinfo{volume}{16} (\bibinfo{year}{2007}) \bibinfo{pages}{2075--2090}.
\bibitem[{{Rayimbaev} and {Tadjimuratov}(2020)}]{RayimbaevPRD2020}
\bibinfo{author}{J.~{Rayimbaev}}, \bibinfo{author}{P.~{Tadjimuratov}},
\newblock \bibinfo{title}{{Can modified gravity silence radio-loud pulsars?}},
\newblock \bibinfo{journal}{Phys. Rev. D} \bibinfo{volume}{102}
  (\bibinfo{year}{2020}) \bibinfo{pages}{024019}.
\bibitem[{{Moffat} and {Toth}(2009{\natexlab{a}})}]{Moffat2009MNRAS}
\bibinfo{author}{J.~W. {Moffat}}, \bibinfo{author}{V.~T. {Toth}},
\newblock \bibinfo{title}{{Modified Gravity and the origin of inertia}},
\newblock \bibinfo{journal}{MNRAS} \bibinfo{volume}{395}
  (\bibinfo{year}{2009}{\natexlab{a}}) \bibinfo{pages}{L25--L28}.
\bibitem[{{Moffat} and {Toth}(2009{\natexlab{b}})}]{Moffat2009MNRASa}
\bibinfo{author}{J.~W. {Moffat}}, \bibinfo{author}{V.~T. {Toth}},
\newblock \bibinfo{title}{{The bending of light and lensing in modified
  gravity}},
\newblock \bibinfo{journal}{MNRAS} \bibinfo{volume}{397}
  (\bibinfo{year}{2009}{\natexlab{b}}) \bibinfo{pages}{1885--1892}.
\bibitem[{{Moffat} and {Toth}(2013)}]{Moffat2013Galaxies}
\bibinfo{author}{J.~{Moffat}}, \bibinfo{author}{V.~{Toth}},
\newblock \bibinfo{title}{{Cosmological Observations in a Modified Theory of
  Gravity (MOG)}},
\newblock \bibinfo{journal}{Galaxies} \bibinfo{volume}{1}
  (\bibinfo{year}{2013}) \bibinfo{pages}{65--82}.
\bibitem[{{Moffat} and {Toth}(2021{\natexlab{a}})}]{Moffat2021EPJC}
\bibinfo{author}{J.~W. {Moffat}}, \bibinfo{author}{V.~T. {Toth}},
\newblock \bibinfo{title}{{The cosmological background and the ``external
  field'' in modified gravity (MOG)}},
\newblock \bibinfo{journal}{European Physical Journal C} \bibinfo{volume}{81}
  (\bibinfo{year}{2021}{\natexlab{a}}) \bibinfo{pages}{836}.
\bibitem[{{Moffat} and {Toth}(2021{\natexlab{b}})}]{Moffat2021Universe}
\bibinfo{author}{J.~W. {Moffat}}, \bibinfo{author}{V.~{Toth}},
\newblock \bibinfo{title}{{Scalar-Tensor-Vector Modified Gravity in Light of
  the Planck 2018 Data}},
\newblock \bibinfo{journal}{Universe} \bibinfo{volume}{7}
  (\bibinfo{year}{2021}{\natexlab{b}}) \bibinfo{pages}{358}.
\bibitem[{{Moffat} and {Rahvar}(2013)}]{Moffat2013MNRAS}
\bibinfo{author}{J.~W. {Moffat}}, \bibinfo{author}{S.~{Rahvar}},
\newblock \bibinfo{title}{{The MOG weak field approximation and observational
  test of galaxy rotation curves}},
\newblock \bibinfo{journal}{MNRAS} \bibinfo{volume}{436} (\bibinfo{year}{2013})
  \bibinfo{pages}{1439--1451}.
\bibitem[{{Moffat} and {Rahvar}(2014)}]{Moffat2014MNRAS}
\bibinfo{author}{J.~W. {Moffat}}, \bibinfo{author}{S.~{Rahvar}},
\newblock \bibinfo{title}{{The MOG weak field approximation - II. Observational
  test of Chandra X-ray clusters}},
\newblock \bibinfo{journal}{MNRAS} \bibinfo{volume}{441} (\bibinfo{year}{2014})
  \bibinfo{pages}{3724--3732}.
\bibitem[{{Moffat}(2015{\natexlab{a}})}]{Moffat2015EPJC}
\bibinfo{author}{J.~W. {Moffat}},
\newblock \bibinfo{title}{{Modified gravity black holes and their observable
  shadows}},
\newblock \bibinfo{journal}{European Physical Journal C} \bibinfo{volume}{75}
  (\bibinfo{year}{2015}{\natexlab{a}}) \bibinfo{pages}{130}.
\bibitem[{{Moffat}(2015{\natexlab{b}})}]{Moffat2015EPJCa}
\bibinfo{author}{J.~W. {Moffat}},
\newblock \bibinfo{title}{{Black holes in modified gravity (MOG)}},
\newblock \bibinfo{journal}{European Physical Journal C} \bibinfo{volume}{75}
  (\bibinfo{year}{2015}{\natexlab{b}}) \bibinfo{pages}{175}.
\bibitem[{{Moffat} and {Toth}(2020)}]{Moffat2020PRD}
\bibinfo{author}{J.~W. {Moffat}}, \bibinfo{author}{V.~T. {Toth}},
\newblock \bibinfo{title}{{Masses and shadows of the black holes Sagittarius A*
  and M87* in modified gravity}},
\newblock \bibinfo{journal}{Phys. Rev. D} \bibinfo{volume}{101}
  (\bibinfo{year}{2020}) \bibinfo{pages}{024014}.
\bibitem[{{Rahvar} and {Moffat}(2019)}]{Rahvar2019MNRAS}
\bibinfo{author}{S.~{Rahvar}}, \bibinfo{author}{J.~W. {Moffat}},
\newblock \bibinfo{title}{{Propagation of electromagnetic waves in MOG:
  gravitational lensing}},
\newblock \bibinfo{journal}{MNRAS} \bibinfo{volume}{482} (\bibinfo{year}{2019})
  \bibinfo{pages}{4514--4518}.
\bibitem[{{Moffat} et~al.(2018){Moffat}, {Rahvar}, and
  {Toth}}]{Moffat2018Galaxies}
\bibinfo{author}{J.~{Moffat}}, \bibinfo{author}{S.~{Rahvar}},
  \bibinfo{author}{V.~{Toth}},
\newblock \bibinfo{title}{{Applying MOG to Lensing: Einstein Rings, Abell 520
  and the Bullet Cluster}},
\newblock \bibinfo{journal}{Galaxies} \bibinfo{volume}{6}
  (\bibinfo{year}{2018}) \bibinfo{pages}{43}.
\bibitem[{{{\"O}vg{\"u}n} et~al.(2019){{\"O}vg{\"u}n}, {Sakall{\i}}, and
  {Saavedra}}]{Ovgun2019AP}
\bibinfo{author}{A.~{{\"O}vg{\"u}n}}, \bibinfo{author}{{\.I}.~{Sakall{\i}}},
  \bibinfo{author}{J.~{Saavedra}},
\newblock \bibinfo{title}{{Weak gravitational lensing by Kerr-MOG black hole
  and Gauss-Bonnet theorem}},
\newblock \bibinfo{journal}{Annals of Physics} \bibinfo{volume}{411}
  (\bibinfo{year}{2019}) \bibinfo{pages}{167978}.
\bibitem[{{Tuleganova} et~al.(2020){Tuleganova}, {Izmailov}, {Karimov},
  {Potapov}, and {Nandi}}]{Tuleganova2020GRG}
\bibinfo{author}{G.~Y. {Tuleganova}}, \bibinfo{author}{R.~N. {Izmailov}},
  \bibinfo{author}{R.~K. {Karimov}}, \bibinfo{author}{A.~A. {Potapov}},
  \bibinfo{author}{K.~K. {Nandi}},
\newblock \bibinfo{title}{{Times of arrival (TOA) of signals in the Kerr-MOG
  black hole}},
\newblock \bibinfo{journal}{General Relativity and Gravitation}
  \bibinfo{volume}{52} (\bibinfo{year}{2020}) \bibinfo{pages}{31}.
\bibitem[{{Izmailov} et~al.(2019){Izmailov}, {Karimov}, {Zhdanov}, and
  {Nandi}}]{Izmailov2019MNRAS}
\bibinfo{author}{R.~N. {Izmailov}}, \bibinfo{author}{R.~K. {Karimov}},
  \bibinfo{author}{E.~R. {Zhdanov}}, \bibinfo{author}{K.~K. {Nandi}},
\newblock \bibinfo{title}{{Modified gravity black hole lensing observables in
  weak and strong field of gravity}},
\newblock \bibinfo{journal}{MNRAS} \bibinfo{volume}{483} (\bibinfo{year}{2019})
  \bibinfo{pages}{3754--3761}.
\bibitem[{{Mureika} et~al.(2016){Mureika}, {Moffat}, and
  {Faizal}}]{Mureika2016PLB}
\bibinfo{author}{J.~R. {Mureika}}, \bibinfo{author}{J.~W. {Moffat}},
  \bibinfo{author}{M.~{Faizal}},
\newblock \bibinfo{title}{{Black hole thermodynamics in MOdified Gravity
  (MOG)}},
\newblock \bibinfo{journal}{Physics Letters B} \bibinfo{volume}{757}
  (\bibinfo{year}{2016}) \bibinfo{pages}{528--536}.
\bibitem[{{Manfredi} et~al.(2017){Manfredi}, {Mureika}, and
  {Moffat}}]{Manfredi2017JPCS}
\bibinfo{author}{L.~{Manfredi}}, \bibinfo{author}{J.~{Mureika}},
  \bibinfo{author}{J.~{Moffat}},
\newblock \bibinfo{title}{{Quasinormal Modes of Static Modified Gravity (MOG)
  Black Holes}},
\newblock in: \bibinfo{booktitle}{Journal of Physics Conference Series}, volume
  \bibinfo{volume}{942} of \textit{\bibinfo{series}{Journal of Physics
  Conference Series}}, p. \bibinfo{pages}{012014}.
\bibitem[{{Manfredi} et~al.(2018){Manfredi}, {Mureika}, and
  {Moffat}}]{Manfredi2018PLB}
\bibinfo{author}{L.~{Manfredi}}, \bibinfo{author}{J.~{Mureika}},
  \bibinfo{author}{J.~{Moffat}},
\newblock \bibinfo{title}{{Quasinormal modes of modified gravity (MOG) black
  holes}},
\newblock \bibinfo{journal}{Physics Letters B} \bibinfo{volume}{779}
  (\bibinfo{year}{2018}) \bibinfo{pages}{492--497}.
\bibitem[{{Manfredi} et~al.(2019){Manfredi}, {Mureika}, and
  {Moffat}}]{Manfredi2019JURP}
\bibinfo{author}{L.~{Manfredi}}, \bibinfo{author}{J.~{Mureika}},
  \bibinfo{author}{J.~{Moffat}},
\newblock \bibinfo{title}{{Quasinormal Modes of Modified Gravity (MOG) Black
  Holes}},
\newblock \bibinfo{journal}{Journal of Undergraduate Reports in Physics}
  \bibinfo{volume}{29} (\bibinfo{year}{2019}) \bibinfo{pages}{100006}.
\bibitem[{{Green} et~al.(2018){Green}, {Moffat}, and {Toth}}]{Green2018PLB}
\bibinfo{author}{M.~A. {Green}}, \bibinfo{author}{J.~W. {Moffat}},
  \bibinfo{author}{V.~T. {Toth}},
\newblock \bibinfo{title}{{Modified gravity (MOG), the speed of gravitational
  radiation and the event GW170817/GRB170817A}},
\newblock \bibinfo{journal}{Physics Letters B} \bibinfo{volume}{780}
  (\bibinfo{year}{2018}) \bibinfo{pages}{300--302}.
\bibitem[{{Wondrak} et~al.(2018){Wondrak}, {Nicolini}, and
  {Moffat}}]{Wondrak2018JCAP}
\bibinfo{author}{M.~F. {Wondrak}}, \bibinfo{author}{P.~{Nicolini}},
  \bibinfo{author}{J.~W. {Moffat}},
\newblock \bibinfo{title}{{Superradiance in modified gravity (MOG)}},
\newblock \bibinfo{journal}{JCAP} \bibinfo{volume}{2018} (\bibinfo{year}{2018})
  \bibinfo{pages}{021}.
\bibitem[{{Qiao} et~al.(2020){Qiao}, {Wang}, {Pan}, and {Jing}}]{Qiao2020EPJC}
\bibinfo{author}{X.~{Qiao}}, \bibinfo{author}{M.~{Wang}},
  \bibinfo{author}{Q.~{Pan}}, \bibinfo{author}{J.~{Jing}},
\newblock \bibinfo{title}{{Kerr-MOG black holes with stationary scalar
  clouds}},
\newblock \bibinfo{journal}{European Physical Journal C} \bibinfo{volume}{80}
  (\bibinfo{year}{2020}) \bibinfo{pages}{509}.
\bibitem[{{Moffat}(2021)}]{Moffat2021JCAP}
\bibinfo{author}{J.~W. {Moffat}},
\newblock \bibinfo{title}{{Modified gravity (MOG), cosmology and black holes}},
\newblock \bibinfo{journal}{JCAP} \bibinfo{volume}{2021} (\bibinfo{year}{2021})
  \bibinfo{pages}{017}.
\bibitem[{{D{\"u}zta{\c{s}}}(2020)}]{Duztas2020EPJC}
\bibinfo{author}{K.~{D{\"u}zta{\c{s}}}},
\newblock \bibinfo{title}{{Overspinning Kerr-MOG black holes by test fields and
  the third law of black hole dynamics}},
\newblock \bibinfo{journal}{European Physical Journal C} \bibinfo{volume}{80}
  (\bibinfo{year}{2020}) \bibinfo{pages}{19}.
\bibitem[{{Moffat}(2016)}]{Moffat2016PLBa}
\bibinfo{author}{J.~W. {Moffat}},
\newblock \bibinfo{title}{{LIGO GW150914 and GW151226 gravitational wave
  detection and generalized gravitation theory (MOG)}},
\newblock \bibinfo{journal}{Physics Letters B} \bibinfo{volume}{763}
  (\bibinfo{year}{2016}) \bibinfo{pages}{427--433}.
\bibitem[{{Moffat} and {Zhoolideh Haghighi}(2017)}]{Moffat2017EPJP}
\bibinfo{author}{J.~W. {Moffat}}, \bibinfo{author}{M.~H. {Zhoolideh Haghighi}},
\newblock \bibinfo{title}{{Modified gravity (MOG) and the Abell 1689 cluster
  acceleration data}},
\newblock \bibinfo{journal}{European Physical Journal Plus}
  \bibinfo{volume}{132} (\bibinfo{year}{2017}) \bibinfo{pages}{417}.
\bibitem[{{Moffat} and {Toth}(2019)}]{Moffat2019MNRAS}
\bibinfo{author}{J.~W. {Moffat}}, \bibinfo{author}{V.~T. {Toth}},
\newblock \bibinfo{title}{{NGC 1052-DF2 and modified gravity (MOG) without dark
  matter}},
\newblock \bibinfo{journal}{MNRAS} \bibinfo{volume}{482} (\bibinfo{year}{2019})
  \bibinfo{pages}{L1--L3}.
\bibitem[{{Green} and {Moffat}(2019)}]{Green2019PDU}
\bibinfo{author}{M.~A. {Green}}, \bibinfo{author}{J.~W. {Moffat}},
\newblock \bibinfo{title}{{Modified Gravity (MOG) fits to observed radial
  acceleration of SPARC galaxies}},
\newblock \bibinfo{journal}{Physics of the Dark Universe} \bibinfo{volume}{25}
  (\bibinfo{year}{2019}) \bibinfo{pages}{100323}.
\bibitem[{{Sharif} and {Shahzadi}(2017)}]{Sharif2017EPJC}
\bibinfo{author}{M.~{Sharif}}, \bibinfo{author}{M.~{Shahzadi}},
\newblock \bibinfo{title}{{Particle dynamics near Kerr-MOG black hole}},
\newblock \bibinfo{journal}{European Physical Journal C} \bibinfo{volume}{77}
  (\bibinfo{year}{2017}) \bibinfo{pages}{363}.
\bibitem[{{Lee} and {Han}(2017)}]{Lee2017EPJC}
\bibinfo{author}{H.-C. {Lee}}, \bibinfo{author}{Y.-J. {Han}},
\newblock \bibinfo{title}{{Innermost stable circular orbit of Kerr-MOG black
  hole}},
\newblock \bibinfo{journal}{European Physical Journal C} \bibinfo{volume}{77}
  (\bibinfo{year}{2017}) \bibinfo{pages}{655}.
\bibitem[{{P{\'e}rez} et~al.(2017){P{\'e}rez}, {Armengol}, and
  {Romero}}]{Perez2017PRD}
\bibinfo{author}{D.~{P{\'e}rez}}, \bibinfo{author}{F.~G.~L. {Armengol}},
  \bibinfo{author}{G.~E. {Romero}},
\newblock \bibinfo{title}{{Accretion disks around black holes in
  scalar-tensor-vector gravity}},
\newblock \bibinfo{journal}{Phys. Rev. D} \bibinfo{volume}{95}
  (\bibinfo{year}{2017}) \bibinfo{pages}{104047}.
\bibitem[{{Pradhan}(2019)}]{Pradhan2019EPJC}
\bibinfo{author}{P.~{Pradhan}},
\newblock \bibinfo{title}{{Study of energy extraction and epicyclic frequencies
  in Kerr-MOG (modified gravity) black hole}},
\newblock \bibinfo{journal}{European Physical Journal C} \bibinfo{volume}{79}
  (\bibinfo{year}{2019}) \bibinfo{pages}{401}.
\bibitem[{{Della Monica} et~al.(2022{\natexlab{a}}){Della Monica}, {de
  Martino}, and {de Laurentis}}]{Monica2022Universe}
\bibinfo{author}{R.~{Della Monica}}, \bibinfo{author}{I.~{de Martino}},
  \bibinfo{author}{M.~{de Laurentis}},
\newblock \bibinfo{title}{{Constraining MOdified Gravity with the S2 Star}},
\newblock \bibinfo{journal}{Universe} \bibinfo{volume}{8}
  (\bibinfo{year}{2022}{\natexlab{a}}) \bibinfo{pages}{137}.
\bibitem[{{Della Monica} et~al.(2022{\natexlab{b}}){Della Monica}, {de
  Martino}, and {de Laurentis}}]{Monica2022MNRAS}
\bibinfo{author}{R.~{Della Monica}}, \bibinfo{author}{I.~{de Martino}},
  \bibinfo{author}{M.~{de Laurentis}},
\newblock \bibinfo{title}{{Orbital precession of the S2 star in
  Scalar-Tensor-Vector Gravity}},
\newblock \bibinfo{journal}{MNRAS} \bibinfo{volume}{510}
  (\bibinfo{year}{2022}{\natexlab{b}}) \bibinfo{pages}{4757--4766}.
\bibitem[{{Turimov}(2022)}]{Turimov2022MNRAS}
\bibinfo{author}{B.~V. {Turimov}},
\newblock \bibinfo{title}{{Comment on ``Orbital precession of the S2 star in
  scalar-tensor-vector gravity''}},
\newblock \bibinfo{journal}{MNRAS} \bibinfo{volume}{516} (\bibinfo{year}{2022})
  \bibinfo{pages}{434--436}.
\bibitem[{{Della Monica} et~al.(2023){Della Monica}, {de Martino}, and {de
  Laurentis}}]{Monica2023MNRAS}
\bibinfo{author}{R.~{Della Monica}}, \bibinfo{author}{I.~{de Martino}},
  \bibinfo{author}{M.~{de Laurentis}},
\newblock \bibinfo{title}{{Response to: Comment on 'Orbital precession of the
  S2 star in scalar-tensor-vector gravity'}},
\newblock \bibinfo{journal}{MNRAS} \bibinfo{volume}{521} (\bibinfo{year}{2023})
  \bibinfo{pages}{474--477}.
\bibitem[{{Kolo{\v{s}}} et~al.(2020){Kolo{\v{s}}}, {Shahzadi}, and
  {Stuchl{\'\i}k}}]{Kolos2020EPJC}
\bibinfo{author}{M.~{Kolo{\v{s}}}}, \bibinfo{author}{M.~{Shahzadi}},
  \bibinfo{author}{Z.~{Stuchl{\'\i}k}},
\newblock \bibinfo{title}{{Quasi-periodic oscillations around Kerr-MOG black
  holes}},
\newblock \bibinfo{journal}{European Physical Journal C} \bibinfo{volume}{80}
  (\bibinfo{year}{2020}) \bibinfo{pages}{133}.
\bibitem[{{Rayimbaev} et~al.(2021){Rayimbaev}, {Tadjimuratov}, {Abdujabbarov},
  {Ahmedov}, and {Khudoyberdieva}}]{Rayimbaev2021Gal}
\bibinfo{author}{J.~{Rayimbaev}}, \bibinfo{author}{P.~{Tadjimuratov}},
  \bibinfo{author}{A.~{Abdujabbarov}}, \bibinfo{author}{B.~{Ahmedov}},
  \bibinfo{author}{M.~{Khudoyberdieva}},
\newblock \bibinfo{title}{{Dynamics of Test Particles and Twin Peaks QPOs
  around Regular Black Holes in Modified Gravity}},
\newblock \bibinfo{journal}{Galaxies} \bibinfo{volume}{9}
  (\bibinfo{year}{2021}) \bibinfo{pages}{75}.
\bibitem[{{Penrose} and {Floyd}(1971)}]{Penrose1971NPhS}
\bibinfo{author}{R.~{Penrose}}, \bibinfo{author}{R.~M. {Floyd}},
\newblock \bibinfo{title}{{Extraction of Rotational Energy from a Black Hole}},
\newblock \bibinfo{journal}{Nature Physical Science} \bibinfo{volume}{229}
  (\bibinfo{year}{1971}) \bibinfo{pages}{177--179}.
\bibitem[{{Dadhich} et~al.(2018){Dadhich}, {Tursunov}, {Ahmedov}, and
  {Stuchl{\'\i}k}}]{Dadhich2018MNRAS}
\bibinfo{author}{N.~{Dadhich}}, \bibinfo{author}{A.~{Tursunov}},
  \bibinfo{author}{B.~{Ahmedov}}, \bibinfo{author}{Z.~{Stuchl{\'\i}k}},
\newblock \bibinfo{title}{{The distinguishing signature of magnetic Penrose
  process}},
\newblock \bibinfo{journal}{MNRAS} \bibinfo{volume}{478} (\bibinfo{year}{2018})
  \bibinfo{pages}{L89--L94}.
\bibitem[{{Bhat} et~al.(1985){Bhat}, {Dhurandhar}, and
  {Dadhich}}]{Bhat1985JApA}
\bibinfo{author}{M.~{Bhat}}, \bibinfo{author}{S.~{Dhurandhar}},
  \bibinfo{author}{N.~{Dadhich}},
\newblock \bibinfo{title}{{Energetics of the Kerr-Newman black hole by the
  Penrose process}},
\newblock \bibinfo{journal}{Journal of Astrophysics and Astronomy}
  \bibinfo{volume}{6} (\bibinfo{year}{1985}) \bibinfo{pages}{85--100}.
\bibitem[{{Stuchl{\'\i}k} et~al.(2021){Stuchl{\'\i}k}, {Kolo{\v{s}}}, and
  {Tursunov}}]{Stuchlik2021Universe}
\bibinfo{author}{Z.~{Stuchl{\'\i}k}}, \bibinfo{author}{M.~{Kolo{\v{s}}}},
  \bibinfo{author}{A.~{Tursunov}},
\newblock \bibinfo{title}{{Penrose Process: Its Variants and Astrophysical
  Applications}},
\newblock \bibinfo{journal}{Universe} \bibinfo{volume}{7}
  (\bibinfo{year}{2021}) \bibinfo{pages}{416}.
\bibitem[{{Tursunov} et~al.(2018{\natexlab{a}}){Tursunov}, {Kolo{\v{s}}}, and
  {Stuchl{\'\i}k}}]{Tursunov2018AN}
\bibinfo{author}{A.~A. {Tursunov}}, \bibinfo{author}{M.~{Kolo{\v{s}}}},
  \bibinfo{author}{Z.~{Stuchl{\'\i}k}},
\newblock \bibinfo{title}{{Orbital widening due to radiation reaction around a
  magnetized black hole}},
\newblock \bibinfo{journal}{Astronomische Nachrichten} \bibinfo{volume}{339}
  (\bibinfo{year}{2018}{\natexlab{a}}) \bibinfo{pages}{341--346}.
\bibitem[{{Tursunov} et~al.(2018{\natexlab{b}}){Tursunov}, {Kolo{\v{s}}},
  {Stuchl{\'\i}k}, and {Gal'tsov}}]{Tursunov2018ApJ}
\bibinfo{author}{A.~{Tursunov}}, \bibinfo{author}{M.~{Kolo{\v{s}}}},
  \bibinfo{author}{Z.~{Stuchl{\'\i}k}}, \bibinfo{author}{D.~V. {Gal'tsov}},
\newblock \bibinfo{title}{{Radiation Reaction of Charged Particles Orbiting a
  Magnetized Schwarzschild Black Hole}},
\newblock \bibinfo{journal}{ApJ} \bibinfo{volume}{861}
  (\bibinfo{year}{2018}{\natexlab{b}}) \bibinfo{pages}{2}.
\bibitem[{{Kolo{\v{s}}} et~al.(2021){Kolo{\v{s}}}, {Tursunov}, and
  {Stuchl{\'\i}k}}]{Kolos2021PRD}
\bibinfo{author}{M.~{Kolo{\v{s}}}}, \bibinfo{author}{A.~{Tursunov}},
  \bibinfo{author}{Z.~{Stuchl{\'\i}k}},
\newblock \bibinfo{title}{{Radiative Penrose process: Energy gain by a single
  radiating charged particle in the ergosphere of rotating black hole}},
\newblock \bibinfo{journal}{Phys. Rev. D} \bibinfo{volume}{103}
  (\bibinfo{year}{2021}) \bibinfo{pages}{024021}.
\bibitem[{{Turimov} et~al.(2022){Turimov}, {Boboqambarova}, {Ahmedov}, and
  {Stuchl{\'\i}k}}]{Turimov2022EPJP}
\bibinfo{author}{B.~{Turimov}}, \bibinfo{author}{M.~{Boboqambarova}},
  \bibinfo{author}{B.~{Ahmedov}}, \bibinfo{author}{Z.~{Stuchl{\'\i}k}},
\newblock \bibinfo{title}{{Distinguishable feature of electric and magnetic
  charged black hole}},
\newblock \bibinfo{journal}{European Physical Journal Plus}
  \bibinfo{volume}{137} (\bibinfo{year}{2022}) \bibinfo{pages}{222}.
\bibitem[{{Tursunov} et~al.(2020){Tursunov}, {Stuchl{\'\i}k}, {Kolo{\v{s}}},
  {Dadhich}, and {Ahmedov}}]{Tursunov2020ApJ}
\bibinfo{author}{A.~{Tursunov}}, \bibinfo{author}{Z.~{Stuchl{\'\i}k}},
  \bibinfo{author}{M.~{Kolo{\v{s}}}}, \bibinfo{author}{N.~{Dadhich}},
  \bibinfo{author}{B.~{Ahmedov}},
\newblock \bibinfo{title}{{Supermassive Black Holes as Possible Sources of
  Ultrahigh-energy Cosmic Rays}},
\newblock \bibinfo{journal}{Astrophys. J} \bibinfo{volume}{895}
  (\bibinfo{year}{2020}) \bibinfo{pages}{14}.
\bibitem[{{Ba{\~n}ados} et~al.(2009){Ba{\~n}ados}, {Silk}, and
  {West}}]{BSW2009PRL}
\bibinfo{author}{M.~{Ba{\~n}ados}}, \bibinfo{author}{J.~{Silk}},
  \bibinfo{author}{S.~M. {West}},
\newblock \bibinfo{title}{{Kerr Black Holes as Particle Accelerators to
  Arbitrarily High Energy}},
\newblock \bibinfo{journal}{Phys. Rev. Lett.} \bibinfo{volume}{103}
  (\bibinfo{year}{2009}) \bibinfo{pages}{111102}.

\end{thebibliography}
\end{document}